\documentclass[aps,prb,twocolumn,groupedaddress]{revtex4}
\usepackage{graphicx}
\usepackage{dcolumn}
\usepackage{bm}
\usepackage{color}
\begin{document}
%
%
\title{Dopant-dependent impact of Mn-site doping on the critical-state manganites: $\bm{R}_{\bm{0.6}}$Sr$_{\bm{0.4}}$MnO$_{\bm{3}}$ ($\bm{R}$=La, Nd, Sm, and Gd)}
%
\author{H. Sakai$^1$, K. Ito$^1$, T. Nishiyama$^1$, X. Z. Yu$^2$, Y. Matsui$^2$, S. Miyasaka$^{1, 3}$, and Y. Tokura$^{1, 4, 5}$}
\affiliation{$^1$Department of Applied Physics, University of Tokyo, Tokyo 113-8656, Japan\\
$^2$Advanced Electron Microscopy Group, Advanced Nano Characterization Center, National Institute for Materials Science (NIMS), Tsukuba 305-0044, Japan\\
$^3$Department of Physics, Osaka University, Toyonaka 560-0043, Japan\\
$^4$Cross-Correlation Materials Research Group (CMRG), ASI, RIKEN, Wako 351-0198, Japan\\
$^5$Multiferroics Project, ERATO, Japan Science and Technology Agency (JST), Tokyo 113-8656, Japan}
%
\begin{abstract}
Versatile features of impurity doping effects on perovskite manganites, $R_{0.6}$Sr$_{0.4}$MnO$_{3}$, have been investigated with varying the doing species as well as the $R$-dependent one-electron bandwidth.
In ferromagnetic-metallic manganites ($R$=La, Nd, and Sm), a few percent of Fe substitution dramatically decreases the ferromagnetic transition temperature, leading to a spin glass insulating state with short-range charge-orbital correlation.
For each $R$ species, the phase diagram as a function of Fe concentration is closely similar to that for $R_{0.6}$Sr$_{0.4}$MnO$_{3}$ obtained by decreasing the ionic radius of $R$ site, indicating that Fe doping in the phase-competing region weakens the ferromagnetic double-exchange interaction, relatively to the charge-orbital ordering instability.
We have also found a contrastive impact of Cr (or Ru) doping on a spin-glass insulating manganite ($R$=Gd).
There, the impurity-induced ferromagnetic magnetization is observed at low temperatures as a consequence of the collapse of the inherent short-range charge-orbital ordering, while Fe doping plays only a minor role.
The observed opposite nature of impurity doping may be attributed to the difference in magnitude of the antiferromagnetic interaction between the doped ions.
\end{abstract}
%
%
\maketitle
%
\section{Introduction}
%
Electronic phase competition often manifests itself in strongly correlated electron systems, which leads to a rich phase diagram with multicriticality.
One typical example is hole-doped perovskite manganites RE$_{1-x}$AE$_{x}$MnO$_3$ (RE being a trivalent rare-earth ion and AE being a divalent alkaline-earth ion), where a charge-orbital-ordered insulating (CO/OO) phase and a ferromagnetic metallic (FM) one keenly compete with each other.
A lot of intriguing phenomena such as colossal magnetoresistance and insulator-metal transition induced by various external stimuli are observed near the bicritical point, where the two phases are almost degenerate in free energy.\cite{Tokura2006review}
%
\par
%
In addition to the external field, quenched disorder arising from the local lattice distortion and/or doped impurities can also significantly modify the electronic structure in the bicritical region.\cite{Dagotto2001Science, Burgy2001PRL, Burgy2004PRL, Murakami2003PRL, Motome2003PRL, Pradhan2007PRL}
A first-order phase transition line separating the two phases completely disappears and instead phase-separated or mixed glassy states on various length- and time-scales are generated depending on the magnitude of the randomness.
Concerning the phase separation caused by strong disorder, effects of impurity doping onto Mn sites have been intensively investigated so far.
Typical examples include an impact of Cr doping in half-doped CO/OO manganites.\cite{Bernabe1997APLa, Raveau1997JSSCa, Katsufuji1999JPSJa, Kimura1999PRLa}.
A few-percent substitutes of Cr easily destroy the long-range CO/OO phase and induce the FM one locally, which results in the phase-separated ground state with both CO/OO and FM clusters randomly distributed.
In contrast to the above, it has recently been found that Fe doping in a FM manganite locating near the bicritical point, such as (La$_{0.7}$Pr$_{0.3}$)$_{0.65}$Ca$_{0.35}$MnO$_{3}$, has also a marked impact; it dramatically suppresses the FM state and instead (short-range) CO/OO correlation evolves down to low temperatures.\cite{Sakai2007PRBa}
In fact, as small as 5\% Fe doping effectively decreases the ferromagnetic transition temperature, $T_{\rm C}$, by $\sim$70\%, while the same amount of Cr or Ga doping only results in a slight decrease in $T_{\rm C}$ by at most $\sim$10\%.
Thus, the disorder effect on the bicritical-state manganite is likely to significantly depend on the impurity species.\cite{Sakai2007PRBa,Hebert2002SSCa}
%
\par
%
In this study, we aim to reveal the comprehensive features of the dopant-dependent Mn-site doping effect by systematically substituting a variety of transition metals for Mn in a critical system of $R_{0.6}$Sr$_{0.4}$MnO$_{3}$ single crystals.
As shown in the inset to Fig. \ref{fig:phase}, this system exhibits the crossover from the FM phase to the spin-glass insulating (SGI) one by decreasing the ionic radius of $R$ site, or equivalently by reducing the effective one-electron bandwidth $W$.
The change in $W$ is here described as a tolerance factor $F$, defined as such that $F\!=\!(r_{\rm A}\!+\!r_{\rm O})/\sqrt{2}(r_{\rm Mn}\!+\!r_{\rm O})$, with $r_{\rm A}$, $r_{\rm Mn}$, and $r_{\rm O}$ being the (averaged) ionic radii of the perovskite A- and B (Mn)-site cations and oxygen, respectively.
Since the bond angel of Mn-O-Mn in the orthorhombic (GdFeO$_3$-type) lattice deviates from 180$^{\circ}$ continuously with $F$, the smaller $F$, or the smaller ionic radius of the $R$ site, results in the smaller $W$.\cite{Imada1998RMP,Torrance1992PRBa}
We can widely control $T_{\rm C}$ for this system from 370 K to 125 K by changing $W$ from $R$=La to Sm, which will provide an ideal arena for investigating the variation of the Fe-doping effects on the different FM correlations.
By measurements of the transport and magnetic properties as well as the electron diffraction patterns, the ability of Fe doping to selectively weaken the FM state is highlighted, which has virtually the same impact as reducing $W$ or $R$ ion size of the parent system near the phase boundary to the SGI phase.
As complementary to this, we have also studied the impurity effect on the SGI system of $R$=Gd, where the short-ranged CO/OO correlation, but no long-range ordering, exists down to low temperatures as the result of the disorder effect due to the large mismatch in ionic size of Gd and Sr.\cite{Tomioka2003PRBa}
Fe doping gives a minimal influence on this system, whereas Cr and Ru doping effectively induces the FM component at low temperatures probably due to the collapse of the short-range CO/OO state.
The possible origin of such a contrastive impurity doping effect is discussed in the light of the strength of the antiferromagnetic coupling between the dopants.
%
\section{Experiment}
%
Single crystals of $R_{0.6}$Sr$_{0.4}$Mn$_{1-y}$Fe$_y$O$_3$ ($R$=La, Nd, and Sm, $0\!\le\!y\!\le\!0.3$) and Gd$_{0.6}$Sr$_{0.4}$Mn$_{1-y}M_{y}$O$_{3}$ ($M$=Fe, Cr, Ru, and Mn, $0\!\le\!y\!\le\!0.2$) were synthesized by a floating zone method.
Powders of La$_2$O$_3$, Nd$_2$O$_3$, Sm$_2$O$_3$, Gd$_2$O$_3$, SrCO$_3$, Mn$_3$O$_4$, $\alpha$-Fe$_2$O$_3$, Cr$_2$O$_3$, and RuO$_2$ mixed in stoichiometric proportions were first calcined at 1050-1100$^\circ$C for 10-20 hours in air.
The mixture was pulverized and again sintered at 1200-1250$^\circ$C for 30-40 hours in air.
The obtained powders were pulverized and then pressed into a rod with $\sim$5 mm in diameter and $\sim$70 mm in length.
The rod was fired at 1350-1400$^\circ$C for 30-40 hours in air.
The crystal growths were performed in an oxygen atmosphere with rotating the feed and seed rods in an opposite direction at the rate of 15-20 rpm.
The growth rate was set at 5-7 mm/h.
Powder x-ray diffraction patterns exhibit that the obtained crystals are of single phase.
Inductively coupled plasma (ICP) spectroscopy has revealed that their composition is equal to the prescribed ratio with the accuracy of $\sim\pm$0.01.
Sr concentration for some of the $R$=La crystals, however, has been found to be lower by at most $\sim$0.05 than that prescribed.
Since the impurity doping effects investigated below would be insensitive to such a small change in hole doping level (in fact, $T_{\rm C}$ is almost constant around $x\!=\!0.3-0.5$ for La$_{1-x}$Sr$_{x}$MnO$_{3}.$\cite{Urushibara1995PRBa}), we here describe Sr concentration for all the crystals as 0.4 for simplicity.
The magnetization and ac susceptibility were measured with a superconducting quantum interference device (low frequencies $<1512$ Hz) and Quantum Design Physical Property Measurement System (higher frequencies).
The resistivity was measured by a conventional four-probe method with electrodes formed by heat-treatment type silver paint.
Electron diffraction patterns were obtained by a transmission electron microscope, Hitachi HF-3000S, equipped with a liquid helium cooling holder.
%
\section{Results and discussion}
%
\subsection{An overview of electronic phase diagrams for Fe-doped manganites}
%
\begin{figure}
\includegraphics[width=8.5cm]{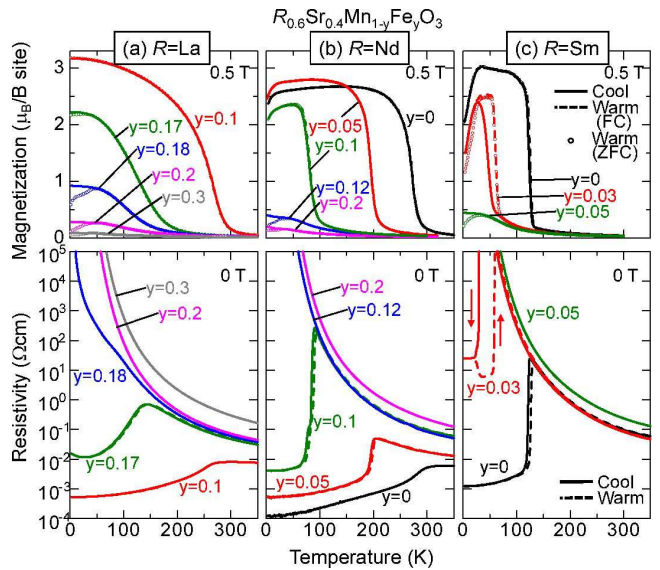}
\caption{\label{fig:RSMFO}(Color online) Temperature profiles of the magnetization at 0.5 T (upper panel) and the resistivity at 0 T (lower panel) for crystals of $R_{0.6}$Sr$_{0.4}$Mn$_{1-y}$Fe$_y$O$_3$ [(a) $R$=La, (b) $R$=Nd, and (c) $R$=Sm, $0\!\le\!y\!\le\!0.3$]. Solid and dashed lines indicate cooling and warming runs, respectively. For the magnetization, the data measured after a zero-field-cooling (ZFC) process are also denoted with the open circles. }
\end{figure}
%
We first show Fe-doping effects on a variety of FM manganites $R_{0.6}$Sr$_{0.4}$MnO$_3$ ($R$=La, Nd, and Sm).
Figure \ref{fig:RSMFO} shows the temperature dependence of the magnetization at 0.5 T (upper panels) and resistivity at 0 T (lower panels) for single crystals of $R_{0.6}$Sr$_{0.4}$Mn$_{1-y}$Fe$_y$O$_3$ [(a) $R$=La, (b) $R$=Nd, and (c) $R$=Sm], where $y$ is the Fe doping level.
The measurements were performed in both cooling and warming runs, denoted as solid and dashed lines respectively.
For the magnetization, the data taken in a warming run after a zero-field-cooling (ZFC) process are shown with open circles while those after a field-cooling (FC) one with dashed lines.
For $R$=La, $T_{\rm C}$ systematically decreases with increasing $y$.
Although the transition to the FM phase is observed for $y\!\le\!0.17$, it disappears and the system remains insulating down to the lowest temperature for $y\!\ge\!0.2$.
For an intermediate Fe concentration of $y$=0.18, the resistivity shows a small anomaly around $\sim$100 K arising from the onset of the FM transition, below which no metallic behavior shows up.
The corresponding magnetization exhibits the clear difference between FC and ZFC below $\sim$50 K, signaling that the system reenters the spin-glass insulating (SGI) phase near this temperature.
With increasing $y$, the SGI phase thus becomes dominant at the ground state by replacing the FM one.
%
\par
%
For $R$=Nd, the Fe doping causes the similar features while its impact is more pronounced.
A rapid decrease in $T_{\rm C}$ accompanied by a steep resistivity drop (by several orders of magnitude) around $T_{\rm C}$ is clearly observed when increasing $y$ up to 0.1.
Such a behavior is typical of FM manganites with a smaller bandwidth, such as $R$=Nd$_{0.5}$Sm$_{0.5}$ or Sm.\cite{Tomioka1997APLa,Saito1999PRBa}
In fact, the features of both resistivity and magnetization for $(R, y)\!=\!(\rm{Nd}, 0.1)$ bear a close resemblance to those for $(R, y)\!=\!(\rm{Sm}, 0)$. [See Fig. \ref{fig:RSMFO}(c)].
Thus, Fe doping effectively reduces the bandwidth of the pristine system by strongly suppressing the FM correlation.
For $R$=Sm, the Fe-doping effect is further emphasized and only 5\% doping is enough to completely destroy the FM phase.
Note that the $y\!=\!0.03$ crystal shows the large residual resistivity of $\sim$25 $\Omega$cm, far above the Ioffe-Regel limit ($\sim$10$^{-3}$ $\Omega$cm).
This indicates that the percolating conduction process will be dominant near the phase boundary to the SGI phase, where the relatively large FM clusters are inhomogeneously distributed in such a low-doping region.
%
\par
%
\begin{figure}
\includegraphics[width=8.5cm]{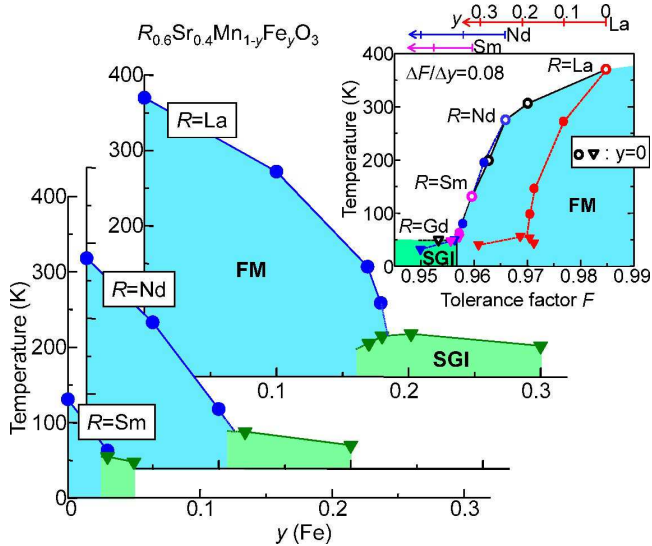}
\caption{\label{fig:phase}(Color online) Phase diagrams for $R_{0.6}$Sr$_{0.4}$Mn$_{1-y}$Fe$_y$O$_3$ ($R$=La, Nd, and Sm) as a function of Fe-doping level $y$. The ferromagnetic metal and spin-glass insulator are denoted by FM and SGI, respectively. The transition temperature to each phase is determined in a warming run. $T_{\rm C}$ and $T_{\rm g}$ are denoted with closed circles and closed triangles, respectively. Inset: phase diagram for $R_{0.6}$Sr$_{0.4}$MnO$_3$ ($R$=La, Nd, Sm, and Gd) as a function of a tolerance factor $F$, which is defined as $F\!=\!(r_{\rm A}\!+\!r_{\rm O})/\sqrt{2}(r_{\rm Mn}\!+\!r_{\rm O})$, where $r_{\rm A}$, $r_{\rm Mn}$, and $r_{\rm O}$ are the (averaged) ionic radii of the perovskite A- and B (Mn)-site cations and oxygen, respectively. Open circles and triangles correspond to $T_{\rm C}$ and $T_{\rm g}$ in this (undoped) system, respectively. The data for $R$=Pr and Nd$_{0.5}$Sm$_{0.5}$ are reproduced from Tomioka $et\ al.$\cite{Tomioka1997APLa} In addition, $T_{\rm C}$ and $T_{\rm g}$ in the Fe-doped systems for each $R$ species are also presented as closed circles and triangles, respectively, with a common relation of $\Delta F\!=\!0.08\Delta y$ shown in the upper abscissa.}
\end{figure}
%
Figure \ref{fig:phase} shows the overall phase diagrams for $R_{0.6}$Sr$_{0.4}$Mn$_{1-y}$Fe$_y$O$_3$ ($R$=La, Nd, and Sm) as a function of Fe concentration $y$.
$T_{\rm C}$ is determined as the temperature where the resistivity curve shows the sudden drop, while the spin glass transition, $T_{\rm g}$, is as the temperature below which the magnetization data shows the history dependence between FC and ZFC.
In all the compounds, a few percent of Fe substitution dramatically destabilizes the FM phase.
$T_{\rm C}$ rapidly decreases down to $\sim$50 K with increasing $y$, and then the FM phase is taken over by the SGI one with almost constant $T_{\rm g}$ ($\sim$50 K).
The value of $y$ necessary to terminate the FM phase systematically decreases from $\sim$0.2 down to $\sim$0.05 when the $R$ ionic radius, i.e. the effective bandwidth of the system, is decreased from La to Sm.
%
\par
%
The resultant phase diagram with respect to $y$ for each $R$ shows close similarity to that for $R_{0.6}$Sr$_{0.4}$MnO$_3$ obtained by changing the $R$ ions to the smaller ones, as shown in the inset to Fig. \ref{fig:phase}.
In this inset, we have plotted $T_{\rm C}$ and $T_{\rm g}$ versus $y$ (closed symbols) obtained in the Fe-doped crystals together with those versus $F$ (open symbols) for the parent compounds of $R_{0.6}$Sr$_{0.4}$MnO$_3$, assuming the common scaling relation that $\Delta F\!=\!0.08\Delta y$, where $\Delta F$ and $\Delta y$ are the effective variations of the respective values.
Using this scaling relation, the data for the Fe-doped $R$=Nd and Sm crystals are consistent with those for the undoped ones as a function of $F$.
For the $R$=La system, however, it does not seem to work well; the values of $T_{\rm C}$ for the Fe-doped systems deviate downward from those for the undoped ones.
Concerning the phase diagram of $R_{0.6}$Sr$_{0.4}$MnO$_3$, $T_{\rm C}$ gradually decreases for $F\!>\!0.965$ with decreasing $F$, while the rate of the decrease in $T_{\rm C}$ is abruptly enhanced for $F\!<\!0.965$, which is anticipated to be due to the rise of the competing instability of the (short-range) charge-orbital ordering.\cite{Tomioka1997APLa,Saito1999PRBa}
The assumed scaling relation is likely to be relevant in the latter, i.e., the phase-competing region, as evidenced for the $R$=Nd and Sm systems.
For the Fe-doped $R$=La system, judging from the value of $T_{\rm C}$, this critical regime will correspond to $y\!\ge\!0.1$, where the rate of the reduction in $T_{\rm C}$ as a function of $y$ agrees fairly well with that for $R$=Nd and Sm.
Therefore, the disagreement of $T_{\rm C}$ versus $y$ and $F$ in the $R$=La system seems to come from the failure of the above scaling relation far from the phase-competing region, where the strong suppression of FM correlation might be represented as a different relation with a larger scaling factor, reflecting the difference in magnitude of the inherent charge-orbital ordering instability.
%
\par
%
In the Fe-doping-induced SGI phase, furthermore, the competing CO/OO correlation survives down to the lowest temperature, as evidenced by the transmission electron microscopy ({\it vide infra}).
This microscopic feature is also analogous to that for the $R$=Gd system, where the CO/OO state remains short-ranged due to the large quenched-disorder effect stemming from the mismatch of the A-site ions.\cite{Tomioka2003PRBa}
%
\subsection{A microscopic nature of Fe-doping-induced spin-glass insulator}
%
\begin{figure}
\includegraphics[width=8.5cm]{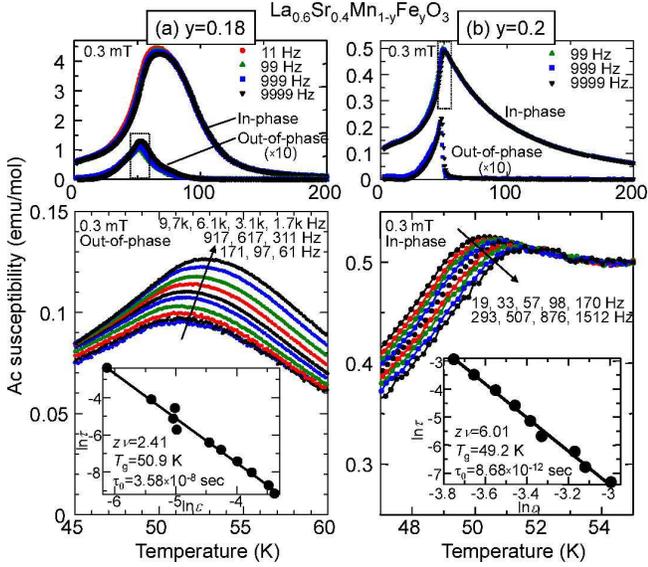}
\caption{\label{fig:ac-chi}(Color online) Upper panels: temperature profiles of the in-phase and out-of-phase components of the ac susceptibility measured at $h\!=\!0.3$ mT with changing the frequency $f$ from 11 Hz to 9999 Hz for crystals of La$_{0.6}$Sr$_{0.4}$Mn$_{1-y}$Fe$_y$O$_3$: (a) $y\!=\!0.18$ and (b) $y\!=\!0.2$. The frequency dependence of the ac susceptibility at $h\!=\!0.3$ mT is studied in more detail in the lower panels: temperature profiles of out-of-phase and in-phase components for (a) $y\!=\!0.18$ and (b) $y\!=\!0.2$, respectively. Inset: dynamical scaling of the observation time $\tau\!=\!1/f$ with the reduced temperature $\varepsilon\!=\!(T_{\rm f}(f)-T_{\rm g})/T_{\rm g}$, where $T_{\rm f}(f)$ and $T_{\rm g}$ denote the frequency-dependent freezing temperature and spin-glass transition temperature, respectively.}
\end{figure}
%
In the vicinity of the phase boundary between FM and SGI, we find the reentrant spin-glass phase, where the SGI transition takes place below $T_{\rm C}$.
The typical example is the $y\!=\!0.18$ crystal for $R$=La.
We show the temperature dependence of the ac susceptibility for this crystal in the upper panel of Fig. \ref{fig:ac-chi}(a).
The in-phase component of the ac susceptibility exhibits a marked increase around 100 K, corresponding to the onset of the FM transition.
Further decreasing the temperature down to $\sim\!50$ K, it shows a shoulder-like anomaly, clearly depending on the frequency.
There, the out-of-phase susceptibility has the maximum, which indicates the reentrance to the SGI phase.\cite{Jonason1996PRBa}
For the $y\!=\!0.2$ crystal, which is SGI locating away from the phase boundary, on the other hand, the susceptibility only shows a cusp structure around 50 K, as presented in the upper panel of Fig. \ref{fig:ac-chi}(b).
The system directly enters the SGI phase without the FM transition.
%
\par
%
To clarify the different nature between these spin-glass phases, we performed a dynamical scaling analysis.\cite{Mathieu2004PRLa, Mathieu2007JPSJa}
The lower panels of Fig. \ref{fig:ac-chi} display the detailed frequency dependence of the ac susceptibility around the spin-glass transition temperature for each crystal.
An observation time $\tau$ is defined as the inverse of the frequency $f$.
From each susceptibility curve, a frequency-dependent freezing temperature $T_{\rm f}(f)$, below which the longest relaxation time of the system exceeds $\tau$, is determined as the peak temperature.
For y=0.18, we employed the out-of-phase component for the scaling analysis, since $T_{\rm f}(f)$ is difficult to assign from the shoulder-like anomaly in the in-phase component.\cite{Jonason1996PRBa}
Using the critical slowing-down power-law relation,\cite{Hohenberg1977RMPa} we have $\tau/\tau_0\!=\!\varepsilon^{-z\nu}$, where $\tau_0$ is the microscopic flipping time for fluctuating entities, $\varepsilon\!=\!(T_{\rm f}(f)-T_{\rm g})/T_{\rm g}$ is the reduced temperature, and $z$ and $\nu$ are the critical exponents.
As shown in the inset to Fig. \ref{fig:ac-chi}, we have obtained a good scaling relation with $\tau_{0}\!=\!3.58\times10^{-8}$ sec, $z\nu\!=\!2.41$, and $T_{\rm g}\!=\!50.9$ K for $y\!=\!0.18$, while $\tau_{0}\!=\!8.68\times10^{-12}$ sec, $z\nu\!=\!6.01$, and $T_{\rm g}\!=\!49.2$ K for $y\!=\!0.2$.
The long flipping time $\tau_{0}$ for $y\!=\!0.18$ will reflect the fluctuation of the FM clusters, which are randomly distributed as a remnant of the high-temperature FM transition.
With further Fe doping, such FM clusters may collapse down to the nanometer scale, leading to the atomic-scale spin-glass phase for $y\!=\!0.2$.
The flipping time for this system, therefore, becomes close to the microscopic spin flipping time ($\sim\!10^{-13}$ s).
The value of the product $z\nu$ is also similar to that of ordinary three-dimensional Heisenberg-like atomic spin glass.\cite{Jonsson2002PRLa}
Thus, the detailed scaling analysis on the spin-glass transition has revealed the clear crossover from the reentrant spin glass with the FM clusters ($cluster$ $glass$) to the atomic-scale spin glass with increasing the Fe-doping level.
%
\par
%
\begin{figure}
\includegraphics[width=4.5cm]{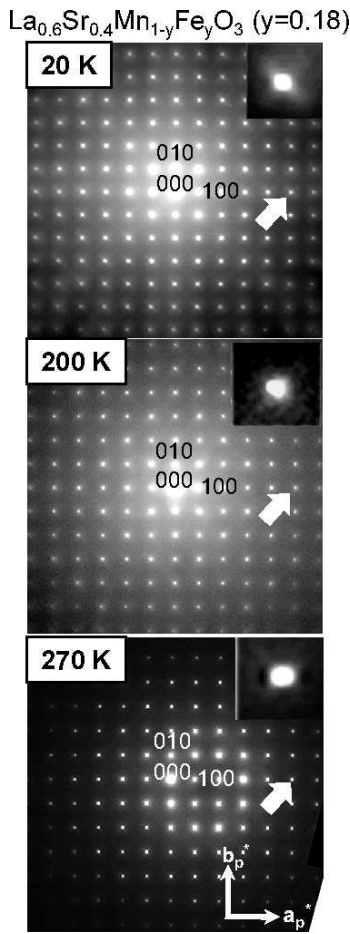}
\caption{\label{fig:diffuse}(Color online) [001]-zone electron diffraction patterns for the  La$_{0.6}$Sr$_{0.4}$Mn$_{0.82}$Fe$_{0.18}$O$_3$ crystal at 20 K (top), 200 K (middle), and 270 K (bottom). The indices are based on the cubic setting. Inset: the magnification of the diffuse scattering around 500 spot (indicated by arrows in the main panels) at each temperature.}
\end{figure}
%
To further investigate the microscopic nature of the Fe-doping-induced SGI phase, we have measured the electron diffraction for the $y\!=\!0.18$ crystal for $R$=La with a transmission electron microscope.
Figure \ref{fig:diffuse} shows the [001]-zone-axis diffraction patterns for this crystal, where the indices are based on the cubic setting.
The most pronounced feature is the appearance of the diffuse scattering around the fundamental Bragg reflections at low temperatures.
As shown in the inset to Fig. \ref{fig:diffuse}, the streak along [110] and [1$\bar{1}$0] directions around 500 reflection evolves upon cooling from room temperature to 20 K.
Such diffuse scattering indicates the presence of the short-range CO/OO correlation, as the remnant of the superlattice reflection of the long-range CO/OO state.
In fact, the similar diffuse scattering (often called Huang scattering) is observed in the $R$=Gd crystal, as was revealed by the x-ray diffraction measurement.\cite{Tomioka2003PRBa}
In the present system, since the FM state is selectively suppressed by Fe doping, the competing CO/OO state will be induced at low temperatures.
Its correlation is, however, only short-ranged and/or dynamical due to the random potential arising from the doped Fe ions.\cite{Sakai2007PRBa}
The reentrant SGI phase for $y\!=\!0.18$, therefore, consists of both the FM and CO/OO components, neither of which is realized as the long-range order even at the ground state.
%
\subsection{Impurity effects on a short-range charge-orbital-ordered manganite}
%
\begin{figure}
\includegraphics[width=6.5cm]{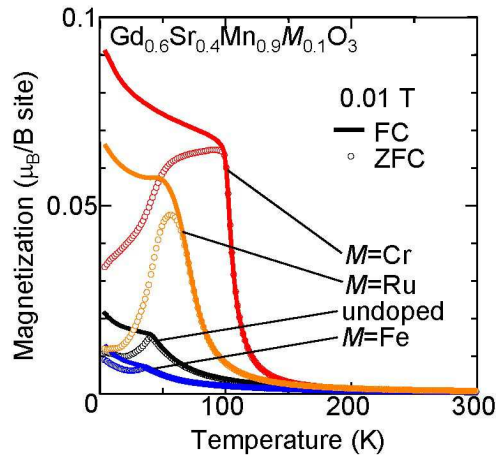}
\caption{\label{fig:GSMMO}(Color online) Temperature profiles of the magnetization at 0.01 T for crystals of Gd$_{0.6}$Sr$_{0.4}$Mn$_{0.9}$$M_{0.1}$O$_3$ [$M$=Fe, Cr, Ru, and Mn (undoped)]. Measurements were performed in a warming run after both ZFC (zero-field cooling; open circles) and FC (field cooling; solid lines) processes.}
\end{figure}
%
We have also investigated the impurity doping effects on the $R$=Gd crystal as a short-range CO/OO system.
Figure \ref{fig:GSMMO} displays the temperature dependence of the magnetization at 0.01 T for single crystals of 10\% $M$-doped Gd$_{0.6}$Sr$_{0.4}$Mn$_{0.9}$$M_{0.1}$O$_3$ ($M$=Fe, Cr, and Ru) and undoped one.
The measurement was performed in a warming run after ZFC (open circles) and FC (solid lines) processes.
The undoped crystal simply shows a cusp structure at $\sim$50 K.
Below this, the distinct difference between ZFC and FC processes is discerned, which is characteristic of the SGI phase.
In the Cr- and Ru-doped crystals, a steep increase in magnetization towards low temperatures is observed, indicating the evolution of the FM state.
As in the case of long-range charge-orbital ordering,\cite{Bernabe1997APLa, Raveau1997JSSCa, Katsufuji1999JPSJa, Kimura1999PRLa, Maignan2001JAPa, Martin2001PRBa} the short-range CO/OO state is anticipated to collapse into the FM one by these kinds of impurities.
Note that the large difference between the FC and ZFC curves is observed below the FM onset temperature.
Such glassy features will reflect that the induced FM state forms only the short-range-ordered clusters.
The Fe-doped crystal, on the other hand, exhibits qualitatively the same magnetization curve as the undoped one, although its size is even reduced nearly by half.
Fe doping thus gives a minimal influence on the SGI phase, i.e., on the short-range CO/OO state, without inducing any FM component.
This tendency also holds for the long-range CO/OO system.\cite{Hebert2002SSCa,Machida2002PRBa}
%
\par
%
\begin{figure}
\includegraphics[width=6.5cm]{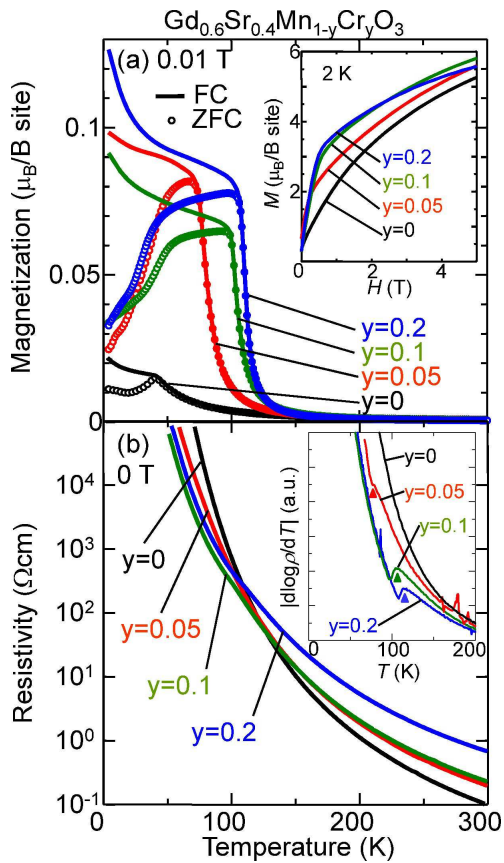}
\caption{\label{fig:GSMCO}(Color online) Temperature profiles of (a) the magnetization at 0.01 T and (b) the resistivity at 0 T in a warming run for crystals of Gd$_{0.6}$Sr$_{0.4}$Mn$_{1-y}$Cr$_{y}$O$_3$ ($0\!\le\!y\!\le\!0.2$). The magnetization data are measured after ZFC and FC processes, which are denoted as open circles and solid lines, respectively. Inset to (a): magnetization $M$ versus magnetic field $H$ up to 5 T at 2 K. Inset to (b): temperature derivative of resistivity $|{\rm d}\log\rho/{\rm d}T|$ versus temperature $T$ at 0 T.}
\end{figure}
%
We further show in Fig. \ref{fig:GSMCO} the detailed results on the Cr concentration dependence of the (a) magnetization and (b) resistivity for single crystals of Gd$_{0.6}$Sr$_{0.4}$Mn$_{1-y}$Cr$_{y}$O$_3$ ($0\!\le\!y\!\le\!0.2$).
Even for $y\!=\!0.05$, a marked evolution of the FM state is observed as an increase in magnetization around 90 K shown in Fig. \ref{fig:GSMCO}(a).
Its onset temperature gradually goes up with increasing $y$, and reaches $\sim$120 K for $y\!=\!0.2$.
The spontaneous magnetization at 2 K also increases with $y$, and amounts to $\sim$2.5 $\mu_{B}$ per formula unit for $y\!=\!0.2$, as displayed in the inset to Fig. \ref{fig:GSMCO}(a).
The corresponding resistivity in Fig. \ref{fig:GSMCO}(b), on the other hand, remains insulating down to the lowest temperature irrespective of $y$.
For $y\!\ge\!0.05$, however, a small anomaly is discerned around the FM onset temperature, which shows up as a clear cusp structure in the temperature derivative of resistivity, as denoted with a closed triangle in the inset to Fig. \ref{fig:GSMCO}(b).
Since below this anomaly temperature the resistivity deviates downward from the expected behavior at higher temperatures, it will stem from the increase in metallic volume fraction induced by Cr doping.
In the long-range CO/OO system like Nd$_{0.5}$Ca$_{0.5}$MnO$_{3}$\cite{Kimura2000PRBa}, as low as 3\% Cr doping is enough to make the system metallic below $T_{\rm C}$ via the percolation of the FM clusters.
In the present system, Cr doping is likely to induce relatively smaller FM clusters in size, reflecting the short-range correlation of the CO/OO state.
Therefore, much higher Cr content is required to produce the enough FM fraction for the percolation path, where the metallic conduction will be inevitably hindered due to the localization effect caused by plenty of Cr disorder. 
This results in the insulating behavior even below the FM onset temperature.
%
\par
%
Let us here discuss the origin of the contrastive nature of Fe and Cr doping, as demonstrated above.
Since both ions prefer to be trivalent state and are antiferromagnetically coupled with Mn ones\cite{Leung1976PRBa,Simopoulos1999PRBa,Studer1999JJAPa}, the carrier density shift or the sign of the superexchange interaction cannot account for their opposite impacts on the bicritical state.
The model previously reported by Martin {\it et al.}\cite{Martin2001PRBa} explains the ability of Cr doping to induce the FM state in the CO/OO structure as follows:
In the CE-type charge-orbital ordering, the FM zigzag chains of Mn$^{3+}$/Mn$^{4+}$ are coupled antiferromagnetically.
The substituted Cr$^{3+}$ will then locate in a zigzag chain with its spin orientation opposite to that of the neighboring zigzag chains.
Furthermore, the neighboring Mn$^{4+}$ on its chain should be antiferromagnetically coupled with Cr$^{3+}$, which consequently reverses all other spins in the FM chain like ``domino effect".
Thus, Cr doping induces the ferromagnetically-coupled (three) zigzag chains, i.e., the FM microclusters.
This scenario may be similarly applicable to the short-range CO/OO system like $R$=Gd.
%
\par
%
In addition to the Cr-Mn coupling, we have to take into consideration the effect of the Cr-Cr interaction when the Cr concentration is further increased.
If the antiferromagnetic interaction between neighboring Cr ions were dominant over other exchange interactions, the induced FM clusters could not expand with aligning their spins in the same direction, which would make the spin system glass-like without developing the FM correlation.
In the present case of Cr doping in the $R$=Gd system, however, the FM correlation continues to grow with increasing the Cr content up to as high as 20\%, as shown in Fig. \ref{fig:GSMCO}.
This means that the Cr-Cr interaction is very weak, or negligible, where the evolution of the FM clusters due to the Cr-Mn interaction can be energetically favored.
For Fe doping in the $R$=Gd crystal, on the other hand, the system never exhibits the onset of the FM transition, as displayed in Fig. \ref{fig:GSMMO}.
The Fe-Fe interaction is so strong that 10\% doping will be enough to suppress the evolution of the induced FM clusters.
In fact, the antiferromagnetic transition temperature of LaFeO$_{3}$ ($T_{\rm N}\!\sim\!750$ K)\cite{Treves1965JAPa, Koehler1960PRa} reaches $\sim$2.6 times as high as that of LaCrO$_{3}$ ($T_{\rm N}\!\sim\!280$ K).\cite{Tezuka1998JSSCa, Sakai1996JSSCa}
We therefore conclude that such a marked difference in the magnitude of the antiferromagnetic interaction between these dopants may be the origin of the contrastive features of Fe and Cr doping.
Note here that the estimated exchange interaction energy $J$ between Fe ions ($\sim$43 K) is merely slightly larger than that between Cr ions ($\sim$37 K), which indicates that the difference in their spin moments may give a dominant contribution to the above tendency.
%
\par
%
\section{CONCLUSION}
%
We have investigated the impurity substitution effect on the electronic structure in $R_{0.6}$Sr$_{0.4}$MnO$_{3}$ ($R$=La, Nd, Sm, and Gd).
This system undergoes a transition from a ferromagnetic metal (for $R$=La, Nd, and Sm) to an insulator with short-range charge-orbital correlation (for $R$=Gd), with changing the $R$-site ionic radius as a control parameter of the one-electron bandwidth.
On the ferromagnetic metallic manganites, Fe doping has a strong impact; a few-percent substitution significantly decreases $T_{\rm C}$.
When the bandwidth is reduced by changing the $R$ site from La to Sm, the Fe content enough to suppress the ferromagnetic transition decreases systematically from 18\% to 5\%.
Further Fe doping makes each system insulating with a spin-glass nature.
In particular, a crossover from reentrant $cluster$ glass to $atomic$ spin glass takes place with increasing the doping level, as revealed by a dynamical scaling analysis of ac susceptibility.
Microscopically, the short-range charge-orbital ordering has been found to coexist in the spin-glass phase by a measurement of transmission electron microscopy.
The resultant phase diagram versus the Fe concentration therefore is closely akin to that of the parent $R_{0.6}$Sr$_{0.4}$MnO$_{3}$ system versus the one-electron bandwidth represented as a tolerance factor.
By selectively suppressing the ferromagnetic metallic correlation, Fe doping is equivalent to the reduction of the bandwidth for the system in the phase-competing region.
In the spin-glass insulating phase for $R$=Gd, on the other hand, Fe doping plays only a minor role while the Cr (or Ru) doping effect is prominent.
In 10-20\% Cr-doped crystals, an evolution of the ferromagnetic state is observed at low temperatures, which will be induced by the collapse of the charge-orbital correlation underlying in the $R$=Gd system.
Reflecting the short-range correlation of the charge-orbital ordering, however, only small-size ferromagnetic clusters seem to be produced, where no metallic conduction manifests itself even below the onset temperature of ferromagnetism.
Such a contrastive tendency of Fe and Cr doping is likely to hold for a variety of bicritical-state manganites, which enables us to bidirectionally control the competing phases by means of the impurity doping.
%
\begin{acknowledgments}
%
We thank Y. Onose and S. Iguchi for fruitful discussions.
We also thank Y. Kiuchi for her help in ICP measurements performed in the Materials Design and Characterization Laboratory, Institute for Solid State Physics, University of Tokyo.
The support to H. S. by the 21st Century COE Program for ``Applied Physics on Strong Correlation'' is appreciated.
This study was partly supported by Grant-in-Aid for JSPS Fellows and for Scientific Research Grants (No. 16076205 and 20340086) from MEXT of Japan.
%
\end{acknowledgments}
%

%

\end{document}